\documentclass[aps,twocolumn,groupedaddress,longbibliography]{revtex4-1}

\usepackage{etex}
\usepackage{amsmath,bm}
\usepackage{courier}
\usepackage[pdftex]{graphicx}
\usepackage{mathtools}
\usepackage[table]{xcolor}
\usepackage{verbatim}
\usepackage{xfrac}

\begin{document}


\title{Active Huygens' Box: Metasurface-Enabled Arbitrary Electromagnetic Wave Generation Inside a Cavity}


\author{Alex M. H. Wong}
\email[]{alex.mh.wong@cityu.edu.hk}
\affiliation{Department of Electronic Engineering, State Key Laboratory of Millimeter Waves, City University of Hong Kong, Hong Kong SAR, China}
\author{George V. Eleftheriades}
\email[]{gelefth@ece.utoronto.ca}
\affiliation{The Edward S. Rogers Sr. Department of Electrical and Computer Engineering, University of Toronto, Canada}


\date{\today}

\begin{abstract}
This work investigates the generation of arbitrary electromagnetic waveforms inside a geometrical area enclosed by an active metasurface. We introduce the contraption of the Huygens' box, where a region of space is enclosed by an active Huygens' metasurface. We show that, upon generating the necessary electric and magnetic currents, we can create any desired electromagnetic field inside the Huygens' box. Using this method, we demonstrate through simulation and experiment the generation of travelling plane waves, a standing plane wave and a Bessel wave inside a metallic cavity. By linear superposition of these unconventional ``modes'', we experimentally demonstrate, for the first time, a subwavelength superoscillation focal spot formed without involving evanescent EM waves, and without an accompanying region of exorbitantly high waveform energy. The Huygens' box brings controlled waveform generation to an unprecedented level, with far-reaching implications to imaging, communication and medical therapy.
\end{abstract}


\maketitle


In the late 17th century, Huygens proposed a theory on light propagation that later became a pillar of classical electromagnetics. His proposal, now known as the Huygens' principle, stipulated that light propagates as a wave: the spatial locations reached by the wave would emanate secondary spherical wavelets that constructively interfere in the direction of light propagation \cite{Huygens1690}. While imperfect, the Huygens' principle was a crucial part of the wave theory of light, which successfully explained light propagation phenomena like reflection, refraction and diffraction off surfaces. Later refinements reconciled the Huygens' principle with the mathematical formulation of electromagnetic waves. Fresnel showed that upon adding an obliquity factor, the Huygens' principle agrees with Kirchhoff's formulation of the scalar theory of electromagnetic waves \cite{Fresnel1866}. Love and Schelkunoff \cite{Love1901,Schelkunoff1936} generalized the Huygens' principle into the electromagnetic equivalence principle, which states that the electromagnetic fields within a region can be generated by any equivalent source in the form of surface currents (typically both electric and magnetic) at the region boundary. Today, the equivalence principle is regarded as a fundamental pillar of classical electromagnetics.

The recent development of electromagnetic metasurfaces \cite{Holloway2005,Yu2011,Kim2014,Estakhri2016,Diaz-Rubio2017,AMHW2018PRX,Aieta2012,Ni2013LSA,Chen2017AdvMater,Grbic2008Science,Markley2008,Rogers2012,Bomzon2002,Pfeiffer2013APL,Selvanayagam2014,Leberer2005,Hunt2013,Epstein2016NatComm,Walther2012,Ni2013,Wen2015} --- artificially engineered surfaces possessing designer electromagnetic effects that are sometimes unfound in nature --- can be understood in the perspective of the Huygens' principle. Essentially, when illuminated by a known incident electromagnetic radiation, the subwavelength features on a metasurface produce the required secondary sources which interfere to generate the desired scattered field. Various demonstrations have showcased the metasurface as a ubiquitous tool which can perform anomalous reflection, refraction and diffraction \cite{Yu2011,Kim2014,Estakhri2016,Diaz-Rubio2017,AMHW2018PRX}, lensing \cite{Aieta2012,Ni2013LSA,Chen2017AdvMater}, subwavelength focusing \cite{Grbic2008Science,Markley2008,Rogers2012}, polarization operation \cite{Bomzon2002,Pfeiffer2013APL,Selvanayagam2014}, antenna beam shaping \cite{Leberer2005,Hunt2013,Epstein2016NatComm}, holography and image formation \cite{Walther2012,Ni2013,Wen2015} among other applications. These works show that, by invoking the Huygens' principle through an artificial planar structure, one can design an electromagnetic wave transformation device that is flexible, low profile, lightweight and low cost compared to its replacement. By extension, one envisions that a greater degree of control can be achieved when a metasurface takes on a curved geometry in 3D space \cite{Diaz-Rubio2018ARXIV}. More importantly, far greater flexibility can be endowed when the equivalent sources are active (impressed) instead of passively induced (secondary). However, with few exceptions \cite{Chen2011PRB,Selvanayagam2012}, this area has heretofore remain unexplored.

This paper demonstrates equivalence-inspired total wavefront control with a metasurface that wraps around a region of interest. We use an active Huygens' metasurface to excite the requisite electric and magnetic currents at the boundary of a square region which we call the Huygens' box, thus achieving unprecedented control of the fields inside the box. We show, through numerical simulation and experimental measurement, the synthesis of a standing wave, travelling waves, a Bessel wave and a subwavelength focus inside the Huygens' box. In particular, through the superposition of travelling waves, one can construct an arbitrary waveform within the Huygens' box. This work demonstrates the direct usage of the electromagnetic equivalence principle to generate a class of waveforms, including waveforms whose existence is originally forbidden in their native environments (in this case a metallic cavity). The Huygens' box allows one to generate and control an arbitrary electromagnetic waveform to unprecedented levels inside an enclosed area, with far-reaching ramifications in imaging, communication and medical therapy.

\section{Results}

\subsection{Electromagnetic Equivalence Principle}

The equivalence principle is a fundamental principle in electromagnetics and states that multiple (equivalent) sources can produce the same electromagnetic field in a region of space \cite{Harrington2001}. One way of generating such an equivalent source is by exciting electric and magnetic currents at the boundary of a region of interest that generate the required electric and magnetic field discontinuities across it. When properly tuned, such an equivalent source boundary electromagnetically separates the regions inside and outside the boundary: each region is then able to assume an electromagnetic field distribution unaffected by the sources and field distribution in the other region.

\begin{figure}[tb]
  \centering
  \includegraphics[width=85mm]{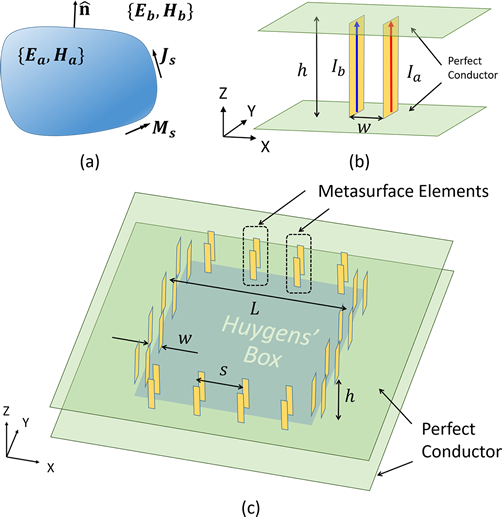}
  \caption{(a) Electromagnetic equivalence theory. A set of currents $\left\{J_s, M_s\right\}$ can generate an electromagnetic field $\left\{ E_a, H_a \right\}$ within an enclosed area without affecting $\left\{ E_b, H_b \right\}$, or adjust the electromagnetic field $\left\{ E_b, H_b \right\}$ outside the area without affecting $\left\{ E_a, H_a \right\}$. (b) The twin current filament that serves as the 2D active Huygens' metasurface element. (c) A schematic of the Huygens' box, showing the 2D environment and the placement of the twin current filaments.
  \label{fig:EquivalenceTheory}}
\end{figure}

We formalize the discussion by referring the reader to Fig. \ref{fig:EquivalenceTheory}a. Here $\left\{\mathbf{E_a}, \mathbf{H_a}\right\}$ and $\left\{\mathbf{E_b}, \mathbf{H_b}\right\}$ represent the waves inside and outside the boundary respectively, $\mathbf{\hat{n}}$  denotes the outward-pointing surface normal, and $\left\{\mathbf{J_s}, \mathbf{M_s}\right\}$ denote the electric and magnetic surface currents, which are given by

\begin{equation}
\label{eq:Th_EquivalenceTheory}
\begin{aligned}
\mathbf{J_s} &= \mathbf{\hat{n}} \times \left( \mathbf{H_b} - \mathbf{H_a} \right) \; , \\
\mathbf{M_s} &= -\mathbf{\hat{n}} \times \left( \mathbf{E_b} - \mathbf{E_a} \right) \; . \\
\end{aligned}
\end{equation}

While the concept of electromagnetic equivalence is a theoretical topic for classical texts in electromagnetics \cite{Harrington2001,Balanis1989}, recent advances in metasurfaces allow one to physically realize such currents and achieve, to an unprecedented level, the arbitrary generation and/or control of an electromagnetic waveform inside an enclosed environment. Specifically, with a passive Huygens' metasurface \cite{Pfeiffer2013PRL,Selvanayagam2013OptEx,Epstein2016JOSAB,Chen2018} one can design the excited electric and magnetic currents along the surface which arise from a predefined incident wave, and thereby achieve one-sided electromagnetic behaviour. In this work, we demonstrate that, by (i) wrapping the metasurface to enclose a region and (ii) using an active Huygens' metasurface, we can directly, actively and arbitrarily tune the electromagnetic fields interior to the metasurface boundary without affecting the electromagnetic wave exterior to the metasurface boundary, and vice versa. This approach can achieve extreme transformations that are not possible with passive and/or planar Huygens' metasurfaces.

\subsection{Huygens' Box Environment}
To simplify the discussion of the conducted simulations and experiments, we consider a 2D, TMz electromagnetic environment where all currents and fields remain invariant in the z-direction and the set of $\left\{H_x, H_y, E_z\right\}$ is non-zero in general. These conditions are met in a parallel-plate waveguide environment which we will discuss later in this paper. In this environment, we define as our region of interest a square boundary of side length $L$ enclosing the area $\left|x\right| \leq L/2$, $\left|y\right| \leq L/2$. We shall call this contraption the Huygens' box: in the following, we report arbitrary waveform generation inside this region through the proper excitation of electric and magnetic currents on its boundary.

\subsection{Active Huygens' Metasurface}
We form a simple active Huygens' metasurface using the twin current filament as the Huygens' source element \cite{AMHW2015APS,AMHW2016MELECON}. As depicted in Fig. \ref{fig:EquivalenceTheory}b, this element consists of two line currents along the z-direction, equidistant from the metasurface boundary, which for illustrative purposes runs along the y-direction. The line source with current $I_a$ is located just inside the Huygens' box, the one with current $I_b$ is located just outside.

The Supplementary Information provides details on the operation mechanism of this active metasurface element. We hereby simply state that the required electric and magnetic currents, as derived from \eqref{eq:Th_EquivalenceTheory}, can be excited by driving the current filaments with

\begin{equation}
\label{eq:Th_ElemCurrents}
\begin{gathered}
I_a = \frac{sJ_s}{2} + j\frac{sM_s}{\omega\mu_0 w} \; , \\
I_b = \frac{sJ_s}{2} - j\frac{sM_s}{\omega\mu_0 w} \; , \\
\end{gathered}
\end{equation}

\noindent where $s$ represents the separation between adjacent metasurface elements, $J_s = \mathbf{J_s} \boldsymbol{\cdot} \mathbf{\hat{z}}$, $M_s = \mathbf{M_s} \boldsymbol{\cdot} \left(\mathbf{\hat{n}} \times \mathbf{\hat{z}}\right)$ and $\boldsymbol{\cdot}$ denotes the inner product. Using \eqref{eq:Th_EquivalenceTheory} and \eqref{eq:Th_ElemCurrents}, one can find the necessary surface currents, then the corresponding excitation currents to the active Huygens' metasurface, that will synthesize or control the electromagnetic waveform in a desired manner.

\subsection{Simulation Results}

\subsubsection{Simulation Setup}
Using our findings above, we synthesize electromagnetic waveforms inside the Huygens' box using the commercial full-wave electromagnetic simulator ANSYS High Frequency Structure Simulator (HFSS). Fig. \ref{fig:EquivalenceTheory}c shows a diagram of the simulation. The simulation is performed at $f=1$ GHz, with a corresponding free-space wavelength of $\lambda=300$ mm. Two metallic conductors, spaced less than half-wavelength apart ($h=44.5\mathrm{mm}=0.148\lambda$), form a parallel-plate environment which enforces z-direction electromagnetic field invariance for all propagating modes. The simulation area is 2$\lambda$ by 2$\lambda$ in the xy-plane, and is terminated by a perfectly matched layer absorbing boundary. A Huygens' box with size $L = \lambda$ is generated in the center of the simulation domain, with element dimensions $s=\lambda/4$, and $w=\lambda/20$. In the following, we show simulation results using line sources as depicted in Fig. \ref{fig:EquivalenceTheory}b, but in practice we find that the current filament can be replaced by a dipole or a monopole. We hypothesize that any small antenna with a matching current direction and a reasonably isotropic radiation pattern would serve the purpose, since the field emanating from such an antenna would couple similarly into the propagating modes of the parallel-plate environment.

\begin{figure*}[tb]
  \centering
  \includegraphics[width=120mm]{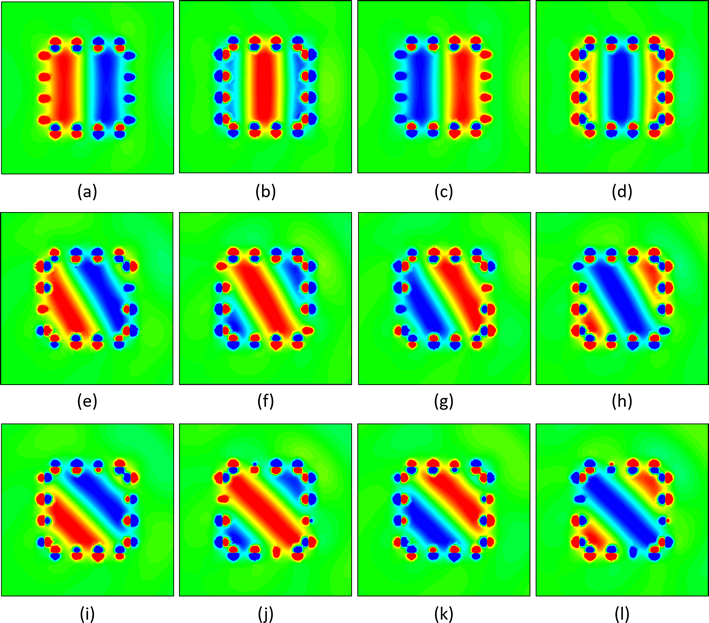}
  \caption{Full-wave simulation results for generating plane waves inside the Huygens' box. Generation of three plane waves, propagating at (a-d) $\theta_1 = 0^\circ$,  (e-h) $\theta_2 = 30^\circ$ and (i-l) $\theta_3 = 45^\circ$ from the x- (horizontal) axis. For each angle of propagation, the instantaneous electric field $E_z(x, y, t)$ is plotted at four different phase points, each spaced $90^\circ$ apart.
  \label{fig:HBSimTW}}
\end{figure*}

\subsubsection{Synthesizing Travelling Waves}
We first demonstrate the generation of travelling plane waves within the Huygens' box. The electric and magnetic fields for a plane wave propagating at an angle $\theta$ with respect to the x-axis can be written as follows:

\begin{equation}
\label{eq:Sim_PWEMFields}
\begin{gathered}
\mathbf{E}(x,y) = E_0 \mathrm{e}^{-j \left(k_x x + k_y y\right) }  \; \mathbf{\hat{z}} \; , \\
\mathbf{H}(x,y) = \frac{E_0}{\eta_0} \mathrm{e}^{-j \left(k_x x + k_y y\right) } \left( \sin \theta \; \mathbf{\hat{x}} - \cos \theta \; \mathbf{\hat{y}} \right) \; , \\
\end{gathered}
\end{equation}

\noindent where $\eta = \sqrt{\mu_0 / \varepsilon_0}$ is the characteristic impedance of free space, $k_x = k_0 \cos{\theta}$, $k_y = k_0 \sin{\theta}$ and $k_0 = 2\pi / \lambda$. In this paper, we assume an $\mathrm{e}^{j\omega t}$ time dependance. To synthesize such a travelling plane wave in the Huygens' box, we find $\left\{\mathbf{E_a}(x,y), \mathbf{H_a}(x,y)\right\}$ by solving \eqref{eq:Sim_PWEMFields} at the metasurface boundary, set $\left\{\mathbf{E_b}, \mathbf{H_b}\right\}$ to zero, then use \eqref{eq:Th_EquivalenceTheory} and \eqref{eq:Th_ElemCurrents} to calculate a set of $\left\{I_a, I_b\right\}$, which form our excitation to the active Huygens' metasurface. Thereafter, we observe the generation of the field profile $\left\{\mathbf{E_a}(x,y), \mathbf{H_a}(x,y)\right\}$ upon applying the excitation currents $\left\{I_a, I_b\right\}$ in the simulation. Fig. \ref{fig:HBSimTW} shows, at four time instants, the time-varying electric field in the simulation environment, corresponding to the generation of three plane waves, with propagation angles of $\theta_1=0^\circ$, $\theta_2=30^\circ$ and $\theta_3=45^\circ$ with respect to the x- (horizontal) axis. From these plots, it is clear that a travelling wave is produced inside the parallel-plate environment at the prescribed angles; it is also clear that the travelling wave is generated only inside the confines of the Huygens' box, even without the presence of a hard boundary (such as a perfect conductor) at the edge of the box to preclude the penetration of the electromagnetic wave. As expected, regions of large field amplitudes exist in the vicinity of the line sources, but the field level decays quickly away from these elements, such that at about an eighth of a wavelength away from the current locations, the travelling plane wave profiles can be clearly observed. By enabling the generation of arbitrary plane waves in an enclosed region, any electromagnetic wave can be synthesized as a superposition of plane waves.

\begin{figure*}[tb]
  \centering
  \includegraphics[width=120mm]{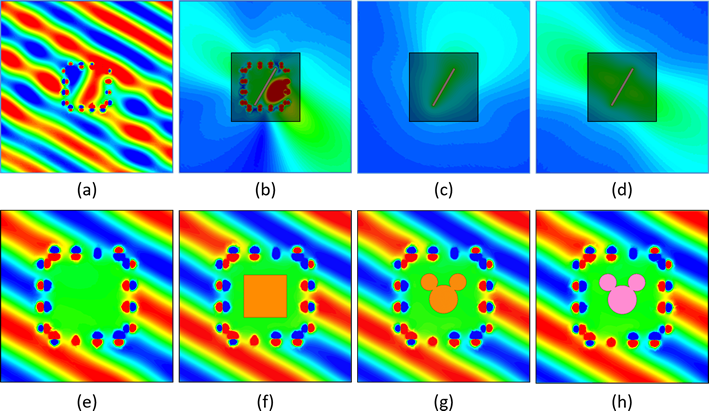}
  \caption{Mimicking and cloaking with the Huygens' box. (a) Electric field distribution when the Huygens' box is excited to turn an incident plane wave at $\theta_i = 60^\circ$ into one propagating at $\theta_{HS} = 150^\circ$ inside the box. (b) Scattering from a metallic plate placed inside the Huygens' box of (a). (c) Scattering from the same metallic plate without the Huygens' box ($\theta_i = 60^\circ$). (d) Scattering from the same metallic plate when the incident wave impinges at $\theta_i = 150^\circ$. (e) Electric field distribution when the Huygens' box is excited to generate a zero-field region upon plane wave incidence at $\theta_i = 60^\circ$. (f-h) The electric field distribution of the Huygens' box in (e), when (f) a metallic box, (g) a collection of metallic cylinders and (h) a collection of ceramic cylinders (Al\textsubscript{2}O\textsubscript{3}, $\varepsilon_r = 9.8$) is placed inside the Huygens' box. The instantaneous electric field is plotted for subfigures (a,e-h); the electric field phasor amplitude is plotted for subfigures (b-d). For subfigures (b-d), the region inside, and immediately outside, the Huygens' box is shaded in grey to emphasize the scattering pattern, which is best seen at some distance away from the Huygens' box.
  \label{fig:HBSimCloak}}
\end{figure*}

\subsubsection{Mimicking and Cloaking Electromagnetic Fields}
The functionality of the Huygens' box is not limited to cases where external fields are absent. Fig. \ref{fig:HBSimCloak}a shows a case where a travelling wave at $\theta_i = 60^\circ$ is twisted $90^\circ$ clockwise within the confines of the box. In this case, the metasurface excitation $\left\{I_a, I_b\right\}$ synthesizes within the Huygens' box the superposition of two plane waves: (i) a travelling wave at $\theta_i=60^\circ$, with the same amplitude but opposite phase to that of the incident wave, and (ii) another travelling wave at $\theta_{HB}=150^\circ$. The first wave cancels the incident electromagnetic field, while the second generates the field profile as seen in Fig. \ref{fig:HBSimCloak}a. Figs. \ref{fig:HBSimCloak}b-d show that, in this situation (Figs. \ref{fig:HBSimCloak}b), an object inside the Huygens' box would assume a scattering pattern that differs from the scattering pattern upon illumination at $\theta_i=60^\circ$ (Fig. \ref{fig:HBSimCloak}c), and approximates the scattering pattering upon illumination at $\theta_{i,eff}=150^\circ$ (Fig. \ref{fig:HBSimCloak}d). Small imperfections lead to slight spurious scattering which causes ripples in Fig. \ref{fig:HBSimCloak}a as well as slight deviations between Fig. \ref{fig:HBSimCloak}b and Fig. \ref{fig:HBSimCloak}d. We expect these deviations would diminish for a larger Huygens' box. Since the waveform inside the Huygens' box can be synthesized at will, the Huygens' box facilitates endless possibilities for mimicking or engineering the scattering profile of any contained object.

Fig. \ref{fig:HBSimCloak}e-h show the special case where the total field vanishes inside the Huygens' box. Because of this absence of electromagnetic fields, the objects inside the box do not scatter, and are effectively cloaked from the external illumination. Fig. \ref{fig:HBSimCloak}f-h show the absence of scattering when a metal box (Fig. \ref{fig:HBSimCloak}f), a cluster of metallic cylinders (Fig. \ref{fig:HBSimCloak}g) and a cluster of ceramic cylinders (Fig. \ref{fig:HBSimCloak}h) are respectively placed inside the Huygens' box. We emphasize that the same set of currents is used in all four situations. This shows that while the currents depend on the cloaking boundary and the external field, they do not change with the size, shape or material of the cloaked object. The cloaking and mimicking properties of an active source boundary was proposed in \cite{Miller2006,Vasquez2009,Zheng2010}, as theoretical works involving scalar (for example, acoustic) waves. In the electromagnetic domain, similar functionalities were investigated under the study of transformation optics \cite{Pendry2006,Lai2009}, and more recently active electromagnetic cloaking \cite{Selvanayagam2012,Selvanayagam2013PRX}. The Huygens' box we hereby propose represents an advancement over previous explorations, in that (i) we have shown that many of the tantalizing possibilities achievable with transformation optics are also achievable with the Huygens' box arrangement, with dramatic improvements in simplicity and compactness compared to the former, and (ii) we have shown, in full-wave simulation analysis, that arbitrary waveform generation, cloaking and scattering engineering is possible with vectorial electromagnetic waves.

\begin{figure}[tb]
  \centering
  \includegraphics[width=85mm]{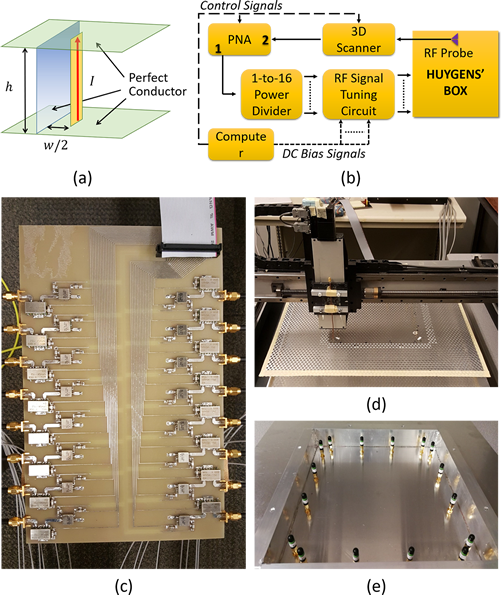}
  \caption{Moving towards an experimental demonstration of the Huygens' box. (a) The mirrored current filament that serves as a simplified 2D active Huygens' metasurface element. (b) A signal flow diagram for the Huygens' Box experiment. Solid, dashed and dotted lines respectively denote RF signals, control signals and DC biases. (c) The 16-channel RF tuning circuit. (d) The 3D scanner and the Huygens' box. (e) The Huygens' box with the top plate removed, showing the box geometry and the monopole antenna elements.
  \label{fig:HBExpApparatus}}
\end{figure}

\subsection{Experimental Results}

\subsubsection{Experimental Setup}
We proceed to experimentally demonstrate the generation of arbitrary waveforms within a Huygens' box. For the experimental demonstration, we adopt a mirrored current filament as our Huygens' metasurface element, which simplifies the twin current filament by dissecting it with a perfect conductor. Fig. \ref{fig:HBExpApparatus}a shows this metasurface element while its operation principle is detailed in the Supplementary Information. The presence of the perfect conductor shorts the electric current, while the required magnetic current can be generated as

\begin{equation}
\label{eq:Sim_ElemCurrent}
I = j\frac{sM_s}{\omega\mu_0 w} \; , \\
\end{equation}

\noindent  and \eqref{eq:Th_EquivalenceTheory} and \eqref{eq:Sim_ElemCurrent} can now be used to synthesize a desired waveform inside the Huygens' box. Our experimental apparatus is shown schematically in Fig. \ref{fig:HBExpApparatus}b; photographs of components the apparatus are shown in Figs. \ref{fig:HBExpApparatus}c-e. An intricate feeding network is designed to drive 16 monopole antennas within the Huygens' box with amplitude and phase-synchronized currents. The currents are fully and independently tunable to achieve active waveform generation. An RF probe measures the electric field within the cavity (Fig. \ref{fig:HBExpApparatus}d) by penetrating through a perforated conductor plate (Fig. \ref{fig:HBExpApparatus}e). The Methods section contains more details on the experimental apparatus.

\begin{figure*}[tb]
  \centering
  \includegraphics[width=120mm]{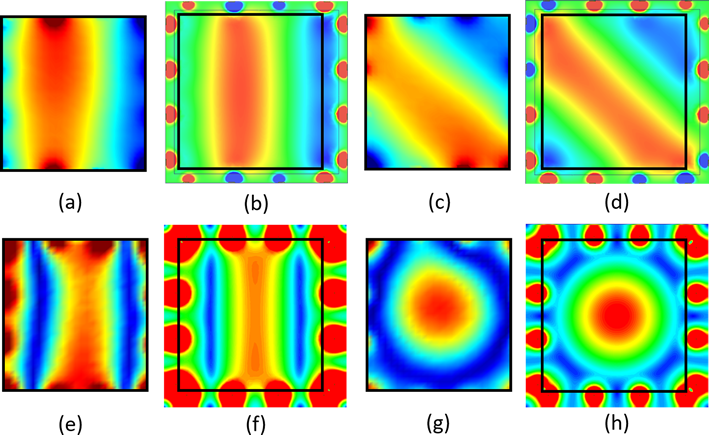}
  \caption{Experimental and full-wave simulation results of electromagnetic wave synthesis with the Huygens' box. The black rectangles indicate the coverage extent of the measurement probe in the experiment. (a-b) The (a) measured and (b) simulated synthesis of a plane wave propagating with $\theta_1 = 0^\circ$. (c-d) The (c) measured and (d) simulated synthesis of a plane wave propagating with $\theta_1 = 45^\circ$. (e-f) The (e) measured and (f) simulated synthesis of a standing wave. (g-h) The (g) measured and (h) simulated synthesis of a zeroth order Bessel function focus. The instantaneous electric field is plotted for subfigures (a-d); the electric field phasor amplitude is plotted for subfigures (e-h).
  \label{fig:HBExp}}
\end{figure*}

\subsubsection{Plane-Wave Excitation}
We first demonstrate the synthesis of a plane wave within the cavity. Figs. \ref{fig:HBExp}a and c show, at one moment in time, the experimentally measured $E_z$ over the cavity for two travelling plane waves, with $\theta_1 = 0^\circ$ and $\theta_2 = 45^\circ$ with respect to the x-axis. Figs. \ref{fig:HBExp}b and d show the corresponding simulation results. As one can observe, good agreement is obtained between experiment and simulation, and both demonstrate the generation of a travelling wave inside our Huygens' box. The superposition of such travelling waves leads to the generation of other interesting waveforms. For example, Figs. \ref{fig:HBExp}e-f show the experimental and simulated generation of a standing wave pattern formed by the equi-amplitude superposition of two travelling waves in the $+$x and $-$x directions. It is of interest to note that the demonstrated waveforms feature very strong electric field components tangential to, and in subwavelength proximity to the metallic cavity walls. Whilst such waves would not exist in a conventional metallic cavity, they can be excited as unconventional ``modes'' when the corresponding excitation is applied to the active metasurface. Further, their combination allows one to form a wide range of waveforms within the Huygens' box.

\subsubsection{Bessel Wave Excitation}
We proceed to synthesize a Bessel function focus within the Huygens' box. The electromagnetic profile for the Bessel focal waveform is \cite{Harrington2001}

\begin{equation}
\label{eq:Exp_BesselWaveform}
\begin{gathered}
\mathbf{E}(r) = E_0 J_0(k_0 r)\; \mathbf{\hat{z}} \; , \\
\mathbf{H}(r) = j\frac{E_0}{\eta_0} J_1(k_0 r) \; \bm{\hat{\phi}} \; , \\
\end{gathered}
\end{equation}

\noindent where $J_n(\cdot)$ represents the $n$'th order Bessel function of the first kind, $r = \sqrt{x^2 + y^2}$ and $\bm{\hat{\phi}} = -\sin \theta \; \mathbf{\hat{x}} + \cos \theta \; \mathbf{\hat{y}}$. One may calculate the necessary Huygens' box excitation by directly substituting \eqref{eq:Exp_BesselWaveform} into \eqref{eq:Th_EquivalenceTheory} and \eqref{eq:Sim_ElemCurrent}. Alternatively, one may invoke the Fourier Bessel transform to express the Bessel waveform as an integral over a continuum of plane waves, which travel in every direction orthogonal to $z$ and constructively interfere at the origin:

\begin{equation}
\label{eq:Exp_BesselIntegral}
\left|\mathbf{E}(r)\right| = E_0 J_0(k_0 r) = E_0 \int_{0}^{2\pi} \mathrm{e}^{-j k_0 \left(x \cos \theta + y \sin \theta \right) } \; d\theta \; . \\
\end{equation}

\noindent We adopt the latter method, then find the excitation currents as a superposition of the excitation currents needed to generate the constituent plane waves.

Fig. \ref{fig:HBExp}g-h demonstrate the measured and simulated electric field profile of a Bessel waveform. Again, simulation and experiment agree to show the successful generation of the Bessel waveform inside the Huygens' box. The Bessel waveform is typically generated as modes in a circular waveguide; further, only a discrete set of Bessel waveforms, whose electric field nulls appear at the waveguide wall, can be generated for a circular waveguide of a specific size. However, with our method, a Bessel waveform with an arbitrary spatial frequency $k_0$ can be synthesized within a rectangular cavity, as long as the corresponding frequency lies within the operation bandwidth of the Huygens' box. Once again this examplifies the versatility of the Huygens' box as a contraption for the generation of arbitrary electromagnetic waveforms.

\subsection{Synthesizing a Superoscillation Focus}

\begin{figure}[tb]
  \centering
  \includegraphics[width=85mm]{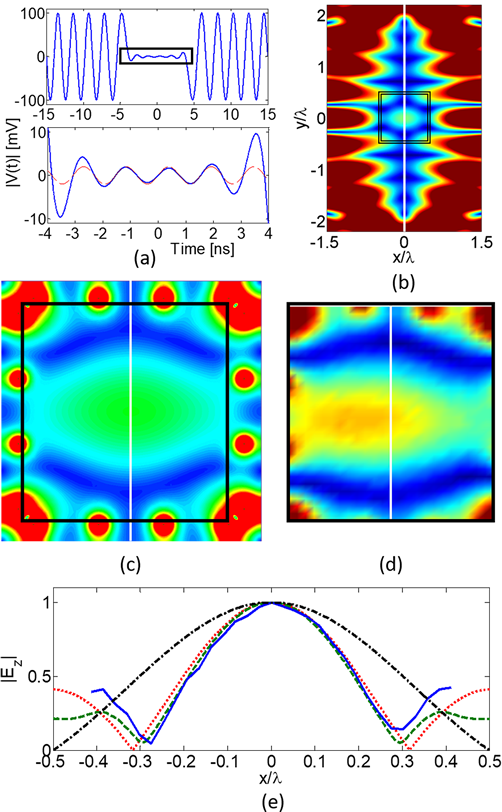}
  \caption{Superoscillation function generation using the Huygens' box. (a) An example of a time domain superoscillation waveform with a bandwidth of 500 MHz (blue, solid). The boxed section in the top panel is shown in a close-up in the bottom panel, and found comparable to a sinusoid oscillating at 650 MHz (red, dashed) (adopted from \cite{AMHW2011TMTT}). (b) A superposition of plane waves leading to a superoscillatory subwavelength focus along the imaging plane (white, solid). The electric field distribution is plotted in a region surrounding the image plane. The proposed Huygens' box location is also depicted (black double-box). (c) The simulated electric field distribution within the Huygens' box. (d) The measured electric field distribution within the Huygens' box. (e) The electric field distribution at the focal plane. The compared waveforms are obtained from theory (red, dotted), simulation (green, dashed) and measurement (blue, solid), and the theoretical 1D diffraction-limited sinc (black, dashdot).
  \label{fig:SuperO}}
\end{figure}

Finally we demonstrate the generation of a subwavelength-focused superoscillation electromagnetic waveform in the Huygens' box. To our knowledge, this waveform represents the first experimental demonstration of a electromagnetic waveform that (i) focuses electromagnetic waves beyond the diffraction limit as discovered by Abb\'{e} and Rayleigh \cite{Abbe1873,Rayleigh1891}, and (ii) achieves this without using evanescent waves and without the accompaniment of a high-energy region \cite{Ferreira2006,Zheludev2008NatMater}. Since only propagating waves are used, subwavelength focusing of this kind avoids the high stored energy issues associated with evanescent waves. This opens the possibility of a variety of applications including hypethermia treatment with localized hot spots in the absence of high energy sidelobes that marred these super-oscillatory focal regions in the past. A more detailed note on superoscillations and sub-diffraction imaging can be found in the Supplementary Information, and in Refs. \cite{Zheludev2008NatMater,AMHW2011TAP,Aharonov2018JOPT}.

\subsubsection{Selective Superoscillation Generation}
Fig. \ref{fig:SuperO}a shows a typical superoscillation waveform, which contains a region in which it oscillates faster than the its highest constituent frequency component, surrounded by
another region which has much higher amplitude \cite{AMHW2011TMTT}. Although the existence of the high-energy region is shown to be unavoidable in a mathematical analysis of waves in free-space \cite{Ferreira2006}, the presence of a metasurface boundary introduces new possibilities for waveform synthesis. In a previous work \cite{AMHW2015SciRep}, we have theoretically shown that with a closed metasurface boundary one can exclude the high-energy region and synthesize only the super-resolution region of a superoscillation waveform. Fig. \ref{fig:SuperO}b shows the electric field profile for a superposition of plane waves that form a 1D subwavelength focus at a prescribed focal plane $x=0$. The electric field profile along this image plane is displayed in Fig. \ref{fig:SuperO}e and compared to the diffraction limit. As one can observe from Fig. \ref{fig:SuperO}e, this waveform is squeezed beyond the diffraction limit at the focal plane. This superoscillation waveform is designed as a combination of plane waves

\begin{equation}
\label{eq:Exp_SuperO}
\begin{gathered}
\mathbf{E}(x,y) = \sum_{n=0}^{N-1} E_n \mathrm{e}^{-jk_{nx} x}\cos\left(k_{ny} y\right) \; \mathbf{\hat{z}} \; , \\
\mathbf{H}(x,y) = \sum_{n=0}^{N-1} \frac{E_n}{\eta_0} \mathrm{e}^{-j k_{nx}x} \left( -j \frac{k_{ny}}{k_0}\sin \left(k_{ny} y \right) \; \mathbf{\hat{x}} \right. \\
\left. - \frac{k_{nx}}{k_0} \cos \left(k_{ny} y \right) \; \mathbf{\hat{y}} \right) \; , \\
\end{gathered}
\end{equation}

\noindent where

\begin{equation}
\label{eq:Exp_kDefn}
\begin{gathered}
k_{nx} = -k_0 + \left(n + \frac{1}{2}\right)\left(\frac{2k_0}{N}\right) \; , \\
k_{ny} = \sqrt{{k_0}^2 - {k_{nx}}^2} \; , \\
\end{gathered}
\end{equation}

\noindent and $\left\{E_n\right\}$ is a set of complex coefficients chosen through a zero-placement method adapted from antenna array design.

Figs. \ref{fig:SuperO}b shows that to generate this waveform in free-space, one must tolerate high-energy regions along the x-axis. Beyond this, it is also typical to find higher energy regions at some longitudinal distance away from the focal plane. To avoid these high-energy regions, we strategically place the Huygens' box (as outlined in Fig. \ref{fig:SuperO}b) to include the superoscillation region only. Fig. \ref{fig:SuperO}c shows the simulated selective generation of this superoscillation waveform in a Huygens' box. As demonstrated here, one can include only the superoscillation region in the Huygens' box and thereby exclude the high-energy regions. Fig. \ref{fig:SuperO}d-e show the experimental measurement which confirms the simulation result and demonstrates the generation of a waveform that is focused beyond the diffraction limit. Again, in departure from earlier works on superoscillation, this is achieved without the generation of a high-energy region. Fig. \ref{fig:SuperO}e plots the waveform distribution along the focal plane and shows good agreement among the theoretical (red, dotted), simulated (green, dashdot) and measured (blue, solid) electric field profiles. All these profiles achieve subwavelength focusing: the spot widths are clearly reduced from the sinc waveform which characterizes the 1D Abb\'{e} diffraction limit (black, dashdot).

\section{Discussion}

In this paper, we examined the synthesis of arbitrary electromagnetic waveforms inside a region enclosed by an active Huygens' metasurface. We introduced an effective 2D environment which we name the Huygens' box, within which we have demonstrated, by calculation, simulation and experimental measurement, the generation of arbitrary electromagnetic waveforms through proper excitation of the enclosing active Huygens' metasurface. We have shown the ability of this simple contraption to perform cloaking and electromagnetic-field mimicking operations. We have also demonstrated the generation of travelling, standing and Bessel waveforms inside a Huygens' box enclosed within a metallic cavity, even though these waveforms cannot exist within a conventional metallic cavity. Finally, we used the Huygens' box to generate a subwavelength focus of propagating waves without exciting the corresponding high-energy region, thus achieving a wave profile very useful for super-resolution electromagnetic focusing and imaging. Though the waveforms are generated in this paper at a design frequency of 1 GHz, experimental measurements (see Supplementary Information) show that the respective waveforms are faithfully synthesized over bandwidth of 11\%, likely limited by the components used in the experiment. Wideband waveform synthesis is very possible given the broadbandedness of the Huygens' metasurface element \cite{AMHW2018PRX}. The Supplementary Information also features a discussion on an extension to a 3D Huygens' box, as well as simulation results showcasing wave generation inside larger Huygens' box with dimensions of several wavelengths. The Huygens' box hereby presented takes electromagnetic waveform control and synthesis to an unprecedented level. We expect it to find promising applications in controlling electric and/or magnetic field profiles in various cloaking, medical imaging and antenna applications which feature the use of electromagnetic waves, in open, enclosed or partially enclosed geometries.

\section*{Methods}
Here we describe our experimental apparatus, which is shown schematically in Fig. \ref{fig:HBExpApparatus}b. The experiment is facilitated by a two-port Agilent programmable network analyzer (PNA). For the purpose of this experiment we use ports 1 and 2 as the transmit and receive ports respectively. A signal at the test frequency is sent from port 1 of the PNA, then evenly split into sixteen portions by a 1-to-16 power divider. The sixteen channels are fed into an RF circuit board shown in Fig. \ref{fig:HBExpApparatus}c. On this circuit board, computer-controlled bias signals regulate the attenuators and phase-shifters to attenuate the amplitude and shift the phase of each RF signal as desired. Hence complete amplitude and phase tuning is afforded to individually control all 16 RF channels. After amplitude and phase adjustment, the 16 RF signals drive 16 monopole antennas located within the Huygens' box, which is shown in Figs. \ref{fig:HBExpApparatus}d-e. As previously explained, monopole antennas function similarly to current filaments with respect to their coupling to propagating modes within the parallel-plate environment. They are chosen in this implementation of the Huygens' box because of their easy installation. Since the mirrored current filament metasurface element features a metallic backing, this effectively turns the Huygens' box into a metallic cavity, with monopole antenna excitations spaced at subwavelength separations. The key device dimensions are: $L = 300$ mm, $s = 75$ mm, $w = 15$ mm and $h = 44.5$ mm. An aluminum plate with hexagonal perforations serves as the top cover of the Huygens' box. The perforations are deeply subwavelength to forbid the leakage of electromagnetic waves, but allow the penetration of a coaxial cable probe which measures the electric field within the Huygens' box. A mechanical scanner scans the measurement probe across an xy-plane located at the vertical center of the Huygens' box. The probe electrically connects to port 2 of the PNA, such that the resultant $S_{21}$ reading on the PNA correlates linearly to a measurement of the z-directed electric field phasor ($E_z$) within the Huygens' box.

\section*{Author Contributions}
The authors jointly conceived the idea behind this research. G.V.E. supervised the work, A.M.H.W. developed the theory, performed the simulation and experiment and wrote the paper.

\end{document}